\documentstyle[preprint,aps,version2]{revtex}
\begin{document}
\draft
\begin{title}
Spin Precession and Time-Reversal Symmetry Breaking\\
in Quantum Transport of Electrons Through Mesoscopic Rings
\end{title}
\author{Ya-Sha Yi, Tie-Zheng Qian, and Zhao-Bin Su}
\begin{instit}
Institute of Theoretical Physics, Chinese Academy of Sciences, 
\\P.O. Box 2735, Beijing 100080, 
People's Republic of China 
\end{instit}
\begin{abstract}
We consider the motion of electrons through a mesoscopic ring in the presence
of spin-orbit interaction, Zeeman coupling, and magnetic flux. 
The coupling between the spin and the orbital degrees of freedom results in
the geometric and the dynamical phases associated with a cyclic evolution
of spin state. Using a non-adiabatic Aharonov-Anandan phase approach, we
obtain the exact solution of the system and identify the geometric and the
dynamical phases for the energy eigenstates. 
Spin precession of electrons encircling the ring can lead to various 
interference phenomena such as oscillating persistent current and
conductance. 
We investigate the transport properties of
the ring connected to current leads to explore  
the roles of the time-reversal symmetry and its breaking therein with
the spin degree of freedom being fully taken into account. 
We derive an exact expression for 
the transmission probability through the ring.
We point out that the time-reversal symmetry breaking due to 
Zeeman coupling can totally invalidate the picture that spin precession
results in effective, spin-dependent Aharonov-Bohm flux for interfering
electrons. Actually, such a picture is only valid in the Aharonov-Casher
effect induced by spin-orbit interaction only.  
Unfortunately, this point has not been realized in prior works on 
the transmission probability in the presence of both SO interaction
and Zeeman coupling. 
We carry out numerical computation to illustrate the joint effects of
spin-orbit interaction, Zeeman coupling and magnetic flux. 
By examining the resonant tunneling of electrons in the weak coupling limit,
we establish a connection between the observable time-reversal symmetry
breaking effects manifested by the persistent current and 
by the transmission probability. For a ring formed by two-dimensional
electron gas, we propose an experiment in which the direction of
the persistent current can be determined by the flux-dependence of
the transmission probability. That experiment also serves to detect if
the electron-electron interaction 
can qualitatively alter the electronic states.
\end{abstract}

\section{Introduction}

The Aharonov-Bohm effect leads to a number of remarkable interference 
phenomena in mesoscopic systems, especially in rings \cite{1}. Based on the 
discovery of the geometric phases \cite{2}, including the 
adiabatic Berry phase \cite{3} and the nonadiabatic 
Aharonov-Anandan (AA) phase \cite{4}, it has been 
predicted that analogous interference phenomena can be induced by the
geometric phases which originate from the interplay between electrons' orbital
and spin degrees of freedom.   
Such interplay can be produced by     
external electric and magnetic fields, which lead to Zeeman
coupling and spin-orbit (SO) interaction respectively.  

Loss {\it et al.} first studied the textured ring embedded in inhomogeneous
magnetic field \cite{5}. They found the inhomogeneity of the field results in 
a Berry phase, which can produce the persistent currents.   
The effects of this Berry phase on conductivity were then discussed \cite{6}.
It was further pointed out that the adiabatic condition is not necessary
for the geometric phase to exist, and the AA phase 
in textured rings can produce the persistent
currents as well \cite{7}.

On the other hand, the Aharonov-Casher (AC) effect \cite{8} in mesoscopic
systems has attracted much attention. Meir {\it et al.} showed for the 
first time that SO interaction in one-dimensional (1D) rings results in  
an effective magnetic flux \cite{9}.
Mathur and Stone then pointed out that observable phenomena induced
by SO interaction are the manifestations of the AC effect in electronic
systems \cite{10}. 
These authors investigated the effects of SO interaction on the                
persistent-current paramagnetism and the quantum transport in disordered
systems, and obtained specific reduction factors for 
harmonics in AB oscillations \cite{9,10,11}.  
In case the AC flux is not random, it can lead to interference
phenomena as AB flux.
Mathur and Stone proposed an observation of the AC oscillation of
the conductance on semiconductor samples \cite{10}.
Balatsky and Altshuler \cite{12} and Choi \cite{13}
studied the persistent currents produced by the AC effect. 

Inspired by the study on textured rings, 
the AC effect has also been analyzed in connection with the
spin geometric phase.
Aronov and Lyanda-Geller considered the spin evolution 
in conducting rings, and found that SO interaction results in a spin-orbit
Berry phase which plays an interesting role in the transmission
probability of the rings \cite{14}. 
In their models, there is a Zeeman coupling from uniform 
magnetic field, but the SO Berry phase can be caused by SO interaction 
alone. So they has indeed shown the existence of the Berry phase in the 
AC effect.
Since SO interaction is usually not strong enough to guarantee the 
validity of adiabatic approximation, a nonadiabatic treatment of the
problem is necessary. In Ref. \cite{15}, we demonstrated the existence of
a nonadiabatic AA phase in the AC effect in 1D rings. We found 
the AC flux and local spin orientations of the electronic eigenstates 
are determined by a spin cyclic evolution. In particular, we showed
the AC phase comprises both the AA and the dynamical phases 
which are acquired in the cyclic evolution, and the adiabatic limit of
the AA phase is just the SO Berry phase. Based on this geometric phase
approach for the AC effect, Oh and Ryu studied the persistent currents 
produced by the cylindrically symmetric SO interaction in 1D rings \cite{16}.

As is well known, SO interaction is time-reversal invariant while 
Zeeman coupling breaks the time-reversal symmetry (TRS).  
Many prior works have shown the significance of the TRS and its breaking 
with regard to various interference phenomena caused by
AB flux and SO interaction. 
It is therefore worthwhile to investigate if the coexistence
of SO interaction and Zeeman coupling can produce any new observable
effect with the spin degree of freedom being fully taken into account. 
However, most of the previous studies have focused on the rings
in the presence of Zeeman coupling or SO interaction only. 
In Ref. \cite{17}, we have demonstrated  
that the competition between Zeeman coupling and SO interaction
can produce persistent currents through the TRS breaking in a 
many-electron ring with a complete set of current-carrying
single-particle states. 
For the transport properties, Aronov and Lyanda-Geller \cite{14} have derived 
a transmission probability for a conducting ring in the presence of
both the SO interaction and the Zeeman coupling, by making use of the 
concept of Berry phase. Unfortunately, they failed to take into account
correctly the different properties of SO interaction and Zeeman coupling under 
the time-reversal transformation. As a result, they have not realized that
their picture of effective flux for the interference of spin-polarized
electrons is actually invalidated by the TRS-breaking Zeeman coupling.
Furthermore, even if the Zeeman coupling is absent, 
their expression for the effective flux induced by the SO interaction 
is still not complete. 
So the transport properties of a ring in the presence of both
SO interaction and Zeeman coupling have not been solved yet and
the roles of TRS and its breaking therein need to be clarified.

In this paper, we will discuss the
transport properties of a ring in the presence of both the SO
interaction and the Zeeman coupling. We will explore the roles of
the TRS and its breaking in the transport phenomena when the spin degree
of freedom is taken into account explicitly. We will also show the connection
between the observable 
TRS-breaking effects manifested by the persistent current 
and by the transmission probability. Throughout the discussion, 
we will emphasize the TRS by investigating how the TRS-breaking
Zeeman coupling affects the thermodynamic and transport properties of 
the system. 
The paper is organized as follows. In Sec. II, we first 
solve the spin cyclic evolution and find the corresponding geometric
and dynamical phases for the system. 
The electronic eigenstates of the closed ring are then derived. 
The geometric phase and the exact solution  
for the ring with only Zeeman coupling
\cite{6,7} or SO interaction \cite{15,16} are shown to be the limit of
zero electric field or the limit of zero magnetic field in our results.
On the other hand, 
the SO Berry phase, first introduced by Aronov and Lyanda-Geller 
\cite{14}, is simply the adiabatic limit of the geometric AA phase here.
In Sec. III, we first derive an exact expression for 
the transfer matrices of the two ring branches (arms) by introducing
four auxiliary spin states, which exhibit the orbital quantum number
dependence of the spin orientations in electronic eigenstates.    
Then we calculate the transmission probability of the ring.
We show that the presence of Zeeman coupling makes the spin orientations
in energy eigenstates depend on the spin and the orbital quantum numbers
simultaneously. 
As a consequence, the effective spin-dependent flux description,  
which has been established for SO interaction only,
becomes inadequate.
That explains why the results in Ref. \cite{14} are not correct. 
When the Zeeman coupling is absent, the derived  
transmission probability agrees with the relations 
obtained by Meir {\it et al.} 
for general spin-independent thermodynamic and transport properties.
We finally carry out some numerical calculations to illustrate the
effects of SO interaction and Zeeman coupling. We find there is 
an interesting and observable correspondence 
between the TRS-breaking effects manifested by the transmission probability 
and by the persistent current. That correspondence, if experimentally
verified or excluded in some specific ring, 
may serve to detect if the electron-electron interaction
is of qualitative importance in determining electronic states.
In Sec. IV, we conclude with a summary of our results.  

\section{Geometric Phase and Exact Solution}
The Hamiltonian for an electron in the electric field 
${\bf E}=-\nabla V$ and the magnetic field ${\bf B}=\nabla\times{\bf A}$ is
\begin{equation}
H=\displaystyle\frac {1}{2m_{e}}({\bf p}-\displaystyle\frac{e}{c}{\bf A})^{2}
+eV-\displaystyle\frac{e\hbar}{4m_{e}^{2}c^{2}}\sigma\!\!\!\!\sigma\cdot{\bf E}
\times ({\bf p}-\displaystyle\frac{e}{c}{\bf A})
-\displaystyle\frac{ge\hbar}{4m_{e}c}
\sigma\!\!\!\!\sigma\cdot{\bf B}.
\end{equation}
We consider a ring that is effectively one-dimensional (1D) and
the fields which are cylindrically symmetric, i.e., 
${\bf E}=E(\cos\chi_{1}{\bf e_{r}}-\sin\chi_{1}{\bf e_{z}})$, 
${\bf B}=B(\sin\chi_{2}{\bf e_{r}}+\cos\chi_{2}{\bf e_{z}})$
in the cylindrical coordinate system. For the ring lying in the
$xy$ plane with its center at the origin, the Hamiltonian is given by
\begin{equation}
H= \displaystyle\frac{\hbar^{2}}{2m_{e}a^{2}}
[-i\displaystyle\frac{\partial}{\partial \theta}
+\phi+\alpha 
(\sin\chi_{1}\sigma_{r}+\cos\chi_{1}
\sigma_{z})]^{2}+\displaystyle\frac{\hbar 
\omega_{B}}{2}(\sin\chi_{2}\sigma_{r}+\cos\chi_{2}\sigma_{z}),
\end{equation}
with $\sigma_{r}=\sigma_{x}\cos\theta 
+\sigma_{y}\sin\theta $,
$\alpha =-\displaystyle\frac{eaE}{4m_{e}c^{2}}$ and
$\omega_{B} =-\displaystyle\frac{geB}{2m_{e}c}$, 
where $a$ is the ring radius, $\theta$ is the angular coordinate
and $\phi$ is the enclosed magnetic flux in unit of flux quantum. 
The eigenvalue equation of the system can be solved through a straightforward 
diagonalization, as presented in Ref. \cite{17}. Here we adopt the 
geometric phase approach \cite{15}, 
in order to identify the geometric and the dynamical phases
in current-carrying eigenstates, which are responsible for transporting 
electrons when the ring is connected to current leads. 
How the phases and the spin orientations jointly affect the transmission
probability will be elaborated in the next section.
 
The cylindrical symmetry of the system leads to the conservation of total
angular momentum
$-i{\partial}/{\partial\theta}
+\frac{1}{2}\sigma_{z}$, which means 
the eigenstates of Hamiltonian (2) are of the form 
$\Psi_{n,\mu}(\theta)=\exp(in\theta)\tilde{\psi}_{n,\mu}(\theta)/\sqrt{2\pi}$, 
in which $\mu=\pm$, $n$ are arbitrary integers, and the spin states are given by
\begin{equation} 
\tilde{\psi}_{n,+}(\theta)=\left[
\begin{array}{c} \cos\displaystyle\frac{\beta_{n}}{2}\\
		 e^{{\it i}\theta}\sin\displaystyle\frac{\beta_{n}}{2}\\
\end{array} \right ];\;\;\;
\tilde{\psi}_{n,-}(\theta)=\left[
\begin{array}{c} \sin\displaystyle\frac{\beta_{n}}{2}\\
		 -e^{{\it i}\theta}\cos\displaystyle\frac{\beta_{n}}{2}\\
\end{array} \right ],
\end{equation}
where $\beta_{n}$ is $\theta$-independent. From $\Psi_{n,\mu}^\dagger
\sigma_{i}\Psi_{n,\mu}$ as a function of $\theta$, it is readily seen that
the local spin orientations at $\theta$ is 
in the direction of $\mu(\cos\beta_{n}{\bf e}_{z}+\sin\beta_{n}{\bf e}_{r})$.
The explicit expression for the spin tilt angle $\beta_{n}$ can be obtained
by introducing a cyclic evolution of spin state 
for electrons encircling the ring, as presented in Ref. \cite{15}. 
The geometric and the dynamical phases associated with the spin precession
can thereby be identified for all of the energy eigenstates to determine
the whole energy spectrum. 
Such an approach has the advantage of explicitly exhibiting 
the geometric phase and the adiabatic criterion, 
which acquire their original meanings in time-dependent problems. 
The spin cyclic evolution is defined by
a time-dependent Schr\"{o}dinger equation
\begin{equation}
{\it i}\hbar\displaystyle\frac{\partial}{\partial t}\psi (t)=H_{s}(t)\psi (t),
\end{equation}
for a spin-$\displaystyle\frac{1}{2}$ particle in a time-varying magnetic field,
where $H_{s}$ is given by
\begin{equation}
\begin{array}{ll}
H_{s}(t)= &
\alpha\hbar\omega[\sin\chi_{1}\cos(\omega t)\sigma_{x}
+\sin\chi_{1}\sin(\omega t)\sigma_{y}+\cos\chi_{1}\sigma_{z}] \\ &
+\displaystyle\frac{1}{2}\hbar\omega_{B}[\sin\chi_{2}\cos(\omega t)\sigma_{x}+
\sin\chi_{2}\sin(\omega t)\sigma_{y}+\cos\chi_{2}\sigma_{z}].
\end{array}
\end{equation}
>From the solution of the cyclic evolution governed by Eq. (4), we obtain
for $\Psi_{n,\mu}$ the spin tilt angle
\begin{equation}
\tan\beta_{n}= 
\displaystyle\frac{2\alpha\omega_{n}\sin\chi_{1}+\omega_{B}\sin\chi_{2}}
{2\alpha\omega_{n}\cos\chi_{1}+\omega_{B}\cos\chi_{2}
-\omega_{n}},
\end{equation}
and the geometric and the dynamical phases  
$\delta_{n,\mu}$ and $\gamma_{n,\mu}$
\begin{equation}
\delta_{n,\mu}
=-\pi(1-\mu\cos\beta_{n}) ,
\end{equation}
\begin{equation}\gamma_{n,\mu}
=-\mu\pi[2\alpha\cos(\beta_{n}-\chi_{1})
+\displaystyle\frac{\omega_{B}}{\omega_{n}}\cos(\beta_{n}-\chi_{2})],
\end{equation}
where $\omega_{n}$ is given by 
$\omega_{0}(n+\displaystyle\frac{1}{2}+\phi)$ with
$\omega_{0}=\displaystyle\frac{\hbar}{ma^{2}}$.
Here the geometric AA phase $\delta_{n,\mu}$ is the -1/2 of the solid angle 
subtended by a circuit traced on a sphere by 
the local spin orientation of $\Psi_{n,\mu}$.
It is readily seen that the Zeeman coupling  makes the spin
orientations of electronic eigenstates depend on the orbital quantum number. 
The consequence of such interplay between the spin and the orbital degrees
of freedom will be explored when we discuss the transport properties of
the ring.  
 
With use of $\beta_{n}$, $\delta_{n,\mu}$, and $\gamma_{n,\mu}$, 
the eigenvalues $E_{n,\mu}$ of $\Psi_{n,\mu}$ is found to be
\begin{equation} 
E_{n,\mu}= \displaystyle\frac{\hbar\omega_{0}}{2}(n+\phi)^{2}
+\displaystyle\frac{\hbar\omega_{0}}{2}(\alpha^{2}-\alpha \cos\chi_{1})
-\displaystyle\frac{\hbar\omega_{n}}{2\pi}
(\delta_{n,\mu}+\gamma_{n,\mu}). 
\end{equation}
The first term in the right side represents the energy from orbital motion,
the second term the zero-point energy, while the third term comes from 
the spin precession originating from
the interplay between the spin and the orbital degrees of freedom.

The exact solution derived above reduces to the various limits that have been 
obtained separately in literatures. The cylindrically symmetric
textured ring, first studied under the adiabatic approximation \cite{5} 
and then exactly solved in Refs. \cite{6,7}, corresponds to the $E=0$ limit.
The AC effect induced by cylindrically symmetric SO interaction in the ring, 
first discussed for vertical field \cite{15} and then investigated for more
general field configurations\cite{16}, 
corresponds to the $B=0$ limit. On the other hand,
with both nonzero $E$ and $B$, the SO Berry phase, first introduced
for a conducting ring \cite{14}, 
is simply the adiabatic limit of the AA phase.

Now we turn to the adiabatic limit of the exact solution. Since the
original stationary Schr{\" o}dinger equation is solved via the solution
of the time-dependent problem, the adiabatic criterion can be easily  
deduced. Comparing $\beta_{n}$ with
the tilt angle of the effective magnetic field in $H_{s}$, we find the 
adiabatic criterion is 
\begin{equation}
\omega_{n}/(2\alpha\omega_{n}\cos\chi_{1}+\omega_{B}\cos\chi_{2})
\to 0.
\end{equation}
For the textured ring with $\alpha=0$, this condition states that the 
Zeeman frequency $\omega_{B}$ must be much larger than the orbital
frequency $\omega_{n}$ unless $\chi_{2}=0$ \cite{6,7}. For the AC effect with
$\omega_{B}=0$, this condition requires that the dimensionless SO
coefficient $\alpha$ 
must be much larger than $1$ unless $\chi_{1}=0$ \cite{15,16}. 

\section{Transmission Probability}

In this section, we discuss the transport properties of 
the ring described by Hamiltonian (2). Now the ring is connected
to external current leads, schematicly illustrated in Fig. 1.
We adopt the standard formulation developed in the study of
quantum oscillations in 1D rings threaded by AB flux \cite{18}.  
In the upper and the lower branches,
the wave amplitudes at one end are related to the wave amplitudes at the
other end by the 
transfer matrices as  
$$\left[\begin{array}{c}\beta_{2} \\ \beta_{2}'\end{array}\right]=\underline
{t}_{I}\left[\begin{array}{c}\beta_{1}' \\ \beta_{1}\end{array}\right],\;\;\;
\left[\begin{array}{c}\gamma_{1} \\ \gamma_{1}'\end{array}\right]=\underline
{t}_{II}'\left[\begin{array}{c}\gamma_{2}' \\ \gamma_{2}\end{array}\right],$$
where $t_{I}$ and $t_{II}'$ denote the transfer matrices of the upper and 
lower branches respectively, and they depend on the energy 
$E$ of the incident wave. At the two junctions, the amplitudes 
of the three outgoing waves $(\alpha ',\beta ',\gamma ')$ are related 
to the amplitudes of the incoming waves $(\alpha ,\beta ,\gamma )$ by
$$\left[\begin{array}{c}\alpha '\\ \beta '\\ \gamma '\end{array}\right]=
\left[\begin{array}{ccc} -(a+b) & \sqrt{\epsilon} & 
\sqrt{\epsilon} \\ \sqrt{\epsilon} & a & b \\ \sqrt{\epsilon} & b & a \end
{array}\right]
\left[\begin{array}{c}\alpha \\ \beta \\ \gamma \end{array}\right],$$
where $a=\pm (\sqrt{1-2\epsilon}-1)/2$ and $b=\pm (\sqrt{1-2\epsilon}+1)/2$
with $0\leq\epsilon\leq 1/2$.
When considering a wave incident from the right junction,
we have $\alpha_{1}^{\dagger}\alpha_{1}
=1$ and $\alpha_{2}=0$. The amplitude of the transmitted wave is 
\begin{equation}
\alpha_{2}'=-\displaystyle\frac{\epsilon}{b^{2}}([b-a\;,\;1]\otimes\sigma_{0})
\underline{t}_{I}
\Pi^{-1}(\left[\begin{array}{c}b-a \\ -1 
\end{array}\right]\otimes\sigma_{0})\alpha_{1},
\end{equation}
with $\Pi$ given by
\begin{equation}
\Pi=\displaystyle\frac{1}{b^{2}}(\left[\begin{array}{cc} b^{2}-a^{2} & a \\
-a & 1 \end{array}\right]\otimes\sigma_{0})\underline{t}_{II}'
(\left[\begin{array}{cc} b^{2}-a^{2} & a \\ -a & 1 \end{array}\right]\otimes
\sigma_{0})\underline{t}_{I}-1\!\! 1,
\end{equation}
where $\sigma_{0}$ is the $2\times 2$ unit matrix in spin space. 
This formulation is in general applicable to the derivation of the 
transmission probability through any ring, provided the
corresponding transfer matrices are known. 
Note that in the study of the ring only threaded by AB flux, 
electrons can be treated as spinless particles, so that all
amplitudes are simply represented by complex numbers and
matrix $\sigma_{0}$ can be dropped. In this paper,
$\alpha_{1}$, $\alpha_{1}'$, $\cdot\cdot\cdot$ 
have to be represented by two-component spinors and 
$\underline{t}_{I}$, $\underline{t}_{II}'$
are $4\times 4$ matrices.

\subsection{Cylindrical symmetry and electronic states in quantum transport}
To derive an explicit expression for the two transfer matrices, 
we first identify the electronic states in the ring 
by making use of its cylindrical symmetry. 
If we write $\underline{t}_{I}$, $\underline{t}_{II}'$ in $2 \times 2$ matrix 
form, then each matrix element is a $2 \times 2$ matrix in spin space.
We can easily conclude that the off-diagonal elements of 
$\underline{t}_{I}$, $\underline{t}_{II}'$ are zero because of the 
conservation of $-i\displaystyle\frac{\partial}{\partial\theta}
+\displaystyle\frac{1}{2}\sigma_{z}$, which indicates that 
in each branch any propagating
wave with fixed energy can posses a well-defined momentum and 
pass each branch without reflection, as a result of the cylindrical symmetry
of the external fields and the absence of scattering potential. So our task
reduces to finding the four $2\times 2$ matrices which respectively relate 
$\beta_{1}'$ with $\beta_{2}$, $\beta_{1}$ with $\beta_{2}'$, 
for the upper branch, and relate $\gamma_{2}'$ with 
$\gamma_{1}$, $\gamma_{2}$ with $\gamma_{1}'$, for the lower branch. These 
four $2\times 2$ matrices are the four nonzero diagonal elements 
of $\underline{t}_{I}$, $\underline{t}_{II}'$ directly.

The electrons' tunneling through the ring is carried out by the energy 
eigenstate of the ring connected to two ideal conductors.
Consider an incident wave with wavevector $k_{F}$.  
The corresponding eigenenergy of the steady transport state 
is $E_{F}={\hbar^{2}k_{F}^{2}}/{2m}$. 
In the right conductor the electronic state is
a superposition of the incident plane wave $\alpha_{1}$ and the reflected 
plane wave $\alpha_{1}'$, while in the left conductor the propagating wave
is just the transmitted plane wave $\alpha_{2}'$. The state inside 
the ring is a superposition of four wavefunctions of energy $E_{F}$. They  
actually determine the four non-zero
matrix elements defined above for the two diagonal transfer matrices.  

To find the four components of the electronic wave inside the ring, we
first use the energy expression 
\begin{equation}
\begin{array}{ll}
\displaystyle\frac{\hbar^{2}k_{F}^{2}}{2m}=E_{n,\mu}= 
& \displaystyle\frac{\hbar\omega_{0}}{2}(n+\phi)^{2}
+\displaystyle\frac{\hbar\omega_{0}}{2}(\alpha^{2}-\alpha \cos\chi_{1})\\
&+\displaystyle\frac{\hbar\omega_{n}}{2}(1-\mu\cos\beta_{n})
+\mu\alpha\hbar\omega_{n}\cos(\beta_{n}-\chi_{1})
+\displaystyle\frac{\mu\hbar\omega_{B}}{2}\cos(\beta_{n}-\chi_{2}). 
\end{array}
\end{equation}
to find four solutions of $n$, 
which are positive $n_{+,+}$ and negative $n_{-,+}$
with $\mu =+$, and positive $n_{+,-}$ and negative $n_{-,-}$ with $\mu=-$. 
For arbitrary $k_{F}$, these quantum numbers are not integers in general. 
For each $n_{\lambda,\mu}$, we can obtain a wavefunction $\Psi_{n_{\lambda,\mu},
\mu}$ which bears the same form as $\Psi_{n,\mu}$ of the closed ring, but 
with $n$ being substituted by $n_{\lambda,\mu}$ and accordingly 
the spin tilt angle $\beta_{n}$ being substituted by $\beta_{n_
{\lambda,\mu}}$ from Eq. (6). These four $\Psi_{n_{\lambda,\mu},\mu}$ 
are actually eigenstates of the Hamiltonian (2) at energy $E_{F}$
but the periodic
boundary condition $\Psi_{n,\mu}(\theta)=\Psi_{n,\mu}(\theta+2\pi)$ is
resolved due to the connection with external conductors.
The electronic wave inside the ring is a superposition of the four 
$\Psi_{n_{\lambda,\mu},\mu}$ by which
the eight amplitudes $\beta_{1},\beta_{1}',\cdot\cdot\cdot$  
can be represented. This is a natural conclusion 
from the steadiness of the electronic state which 
transports electrons at fixed energy $E_{F}$ through the ring. 
With this understanding, we can derive the transfer
matrices in terms of $\Psi_{n_{\lambda,\mu},\mu}$. 

\subsection{Transfer matrices represented by nonorthogonal spin states} 
As shown in Sec. II, the Zeeman coupling brings the dependence on
orbital quantum number to spin orientations. As a result, 
$\Psi_{n_{\lambda,+},+}$ and 
$\Psi_{n_{\lambda,-},-}$, which carry the clockwise ($\lambda=-$) or
the anticlockwise $(\lambda=+)$ wave, are of nonorthogonal spin states
$\tilde{\psi}_{n_{\lambda,-},-}(\theta)$ and
$\tilde{\psi}_{n_{\lambda,+},+}(\theta)$
unless in the absence of Zeeman coupling.
To derive the transfer matrix associated with spin-polarized transport, 
it is crucial to  
distinguish the $\mu=+$ from the $\mu=-$ contribution for 
any wave propagating in fixed direction. 
For this purpose,  
we define four auxiliary spin states 
\begin{equation}
\tilde{\eta}_{\lambda,\mu}(\theta) 
=\displaystyle\frac{1}{R_{\lambda}}(\tilde{\psi}_{n_{\lambda,\mu},\mu}(\theta)
-\tilde{\psi}_{n_{\lambda,-\mu},-\mu}^{\dagger}(\theta)
\tilde{\psi}_{n_{\lambda,\mu},\mu}
(\theta)\tilde{\psi}_{n_{\lambda,-\mu},-\mu}(\theta)), 
\end{equation}
where $R_{\lambda}=1-|\tilde{\psi}_{n_{\lambda,\mu},\mu}^{\dagger}(\theta)
\tilde{\psi}_{n_{\lambda},-\mu}(\theta)|^{2}$. It is easy to verify the 
relations of redefined orthogonality and completeness, 
\begin{equation}
\tilde{\eta}_{\lambda,\mu}(\theta)^{\dagger}
\tilde{\psi}_{n_{\lambda,\nu},\nu}(\theta)=\delta_{\mu\nu}, 
\end{equation}
and
\begin{equation}
\sum_{\mu} 
\tilde{\psi}_{n_{\lambda,\mu},\mu}(\theta)
\tilde{\eta}_{\lambda,\mu}(\theta)^{\dagger}=\sigma_{0}. 
\end{equation}
In the upper branch,
the wave propagating anticlockwisely consists of the two
components $\Psi_{n_{+,+},+}$ and $\Psi_{n_{+,-},-}$. $\beta_{1}'$
and $\beta_{2}$ can thereby be expressed as
\begin{equation}\begin{array}{ll}
\beta_{1}^{'} &=c_{1}\Psi_{n_{+,+},+}(0)+c_{2}\Psi_{n_{+,-},-}(0); \\
\beta_{2} &=c_{1}\Psi_{n_{+,+},+}(\pi)+c_{2}\Psi_{n_{+,-},-}(\pi),
\end{array}
\end{equation}
where $c_1$ and $c_2$ are two specific constants. 
Using Eqs. (3) and (15), we obtain 
\begin{equation}
\beta_{2}=[e^{in_{+,+}\pi}\tilde{\psi}_{n_{+,+},+}(\pi)
\tilde{\eta}_{+,+}^{\dagger}(0)+
e^{in_{+,-}\pi}\tilde{\psi}_{n_{+,-},-}(\pi)\tilde{\eta}_{+,-}^{\dagger}(0)]
\beta_{1}^{'},
\end{equation}
and therefore find the $2\times 2$ matrix which is the first diagonal 
element of $\underline{t}_{I}$. The other three matrix elements in diagonal
$\underline{t}_{I}$ and $\underline{t}_{II}$ can be derived in the same way. 
We finally obtain the two transfer matrices in the form of  
\begin{equation}
\underline{t}_{I}=
\left [ \begin{array}{cc}
\sum_{\mu}e^{in_{+,\mu}\pi}\tilde{\psi}_{n_{+,\mu},\mu}(\pi)
\tilde{\eta}_{+,\mu}^{\dagger}(0) & 0 \\
0 & \sum_{\mu}e^{in_{-,\mu}\pi}\tilde{\psi}_{n_{-,\mu},\mu}(\pi)
\tilde{\eta}_{-,\mu}^{\dagger}(0)
\end{array} \right ];
\end{equation}
\begin{equation}
\underline{t}_{II}^{'}=
\left [ \begin{array}{cc}
\sum_{\mu}e^{in_{+,\mu}\pi}\tilde{\psi}_{n_{+,\mu},\mu}(0)
\tilde{\eta}_{+,\mu}^{\dagger}(\pi) & 0 \\
0 & \sum_{\mu}e^{in_{-,\mu}\pi}\tilde{\psi}_{n_{-,\mu},\mu}(0)
\tilde{\eta}_{-,\mu}^{\dagger}(\pi)
\end{array} \right ].
\end{equation}
>From Eq. (11), the transmission probability for unpolarized incident electrons
is $<\alpha_{2}'^\dagger\alpha_{2}'>$ in which $<\cdot\cdot\cdot>$ denotes
an averaging over $\alpha_{1}$ with fixed $\alpha_{1}^\dagger\alpha_{1}=1$. 
Explicitly, it is given by 
\begin{equation}T=\displaystyle\frac{1}{2}\sum_{ij}\left| 
[-\displaystyle\frac{\epsilon}{b^{2}}([b-a\;,\;1]\otimes\sigma_{0})
\underline{t}_{I}
\Pi^{-1}(\left[\begin{array}{c}b-a \\ -1 
\end{array}\right]\otimes\sigma_{0})]_{ij}\right| ^2,
\end{equation}
in which $\underline{t}_{I}$, $\underline{t}_{II}'$, and $\Pi$ are all known.

\subsection{Success and breakdown of effective flux description} 
In the absence of Zeeman coupling, the 
expression of the transmission probability can be greatly simplified and
be explicitly related to the spin-independent transmission probability
through the ring threaded by AB flux only.
>From Eq. (6), it is obvious that if $\omega_{B}=0$, 
$\tilde{\psi}_{n_{\lambda,\mu},\mu}$ 
are independent of $n_{\lambda,\mu}$ defined
in Eq. (13) and can be denoted by $\tilde{\psi}_{\mu}$ with
$\tilde{\eta}_{\lambda,\mu}=\tilde{\psi}_{\mu}$. 
Combining this fact with Eqs. (15), (16), and (17), 
we see that in the absence of Zeeman coupling, the electronic 
wave in the ring actually consists of two orthogonal amplitudes, which
propagate coherently and independently, with their local spin states being 
given by $\tilde{\psi}_{\mu}$. 
We then turn to the phase shift for spin-polarized electrons.
When $k_{F}a$ is very large and quasiclassical approximation is therefore
applicable, it is worthwhile to write $n_{\lambda,\mu}$ in Eq. (13) as
\begin{equation}
n_{\lambda,\mu}=\lambda k_{F}a-\phi-\displaystyle\frac{1}{2}
(1-\mu \cos\chi_{n_{\lambda,\mu}})
-\mu\alpha\cos\chi_{n_{\lambda,\mu}},
\end{equation}
where the last three terms in the right side are $\displaystyle\frac{1}{2\pi}$
of the AB phase, the spin AA, and the dynamical
phases contributed by the SO interaction, respectively.
In case the Zeeman coupling is absent, the last two terms give the 
$1/2\pi$ of the AC phase, $\Phi_{AC}^{\mu}/2\pi$ \cite{15}, and 
\begin{equation}
n_{\lambda,\mu}=\lambda k_{F}a-\phi+\displaystyle\frac{1}{2\pi}
\Phi_{AC}^{\mu}
\end{equation} 
becomes an exact relation without quasiclassical
approximation. Since $\Phi_{AC}^{\mu}$ is $n$-independent, 
Eq. (23) indicates that the effect of the SO interaction can be regarded as 
an AB effect  
of the effective flux $-\Phi_{AC}^{\mu}/2\pi$ in unit of
$\Phi_{0}$ for the locally polarized electron gases with local spin states
$\tilde{\psi}_{\mu}$. 

We can derive for $\omega_{B}=0$ the transmitted amplitude 
\begin{equation}
\alpha_{2}'(\phi,\alpha_{1})=
\sum_{\mu}[\tilde{\psi}_{\mu}^{\dagger}(0)\alpha_{1}]
t(\phi-\displaystyle\frac{\Phi_{AC}^{\mu}}{2\pi})
\tilde{\psi}_{\mu}(\pi),
\end{equation}
where $t(\phi)$ is the transmitted amplitude for the ring threaded by magnetic
flux $\phi$ \cite{18} with vanishing SO interaction. 
Eq. (24) indicates clearly that
the real electronic wave in the ring is a superposition of the two locally
polarized waves, which enclose different effective fluxes and propagate 
independently. The transmission probability  $T_{AB,AC}$ is given by 
${\alpha_{2}'}^{\dagger}\alpha_{2}'$:
\begin{equation}
T_{AB,AC}(\phi,\alpha_{1})
=\sum_{\mu}|\tilde{\psi}_{\mu}^{\dagger}(0)\alpha_{1}|^{2}T_{AB}(\phi-
\displaystyle\frac{\Phi_{AC}^\mu}{2\pi}),
\end{equation}
where $T_{AB}(\phi)=t^{\dagger}t$ is the transmission probability of the ring 
threaded by magnetic flux $\phi$ with vanishing SO interaction.
To see what happens for unpolarized incident wave, we average $T_{AB,AC}$ over
$\alpha_{1}$ and obtain 
$\bar{T}_{AB,AC}=\sum_{\mu}T_{AB}(\phi-\Phi_{AC}^\mu/2\pi)/2$,
which agrees with the relation predicted in Ref. \cite{9} for 
general spin-independent thermodynamic and transport quantities.

In the competition with the SO interaction, 
the Zeeman coupling brings the $n$-dependence
to the spin orientations of energy eigenstates. The $n$-dependent
spin precession then results in the $n$-dependent spin phases. 
It is seen that in the presence of the Zeeman coupling, the last two terms in
Eq. (22) are $n$-dependent and the effect of the spin phases can no longer
be regarded as that from the effective flux which must be independent
of the specific orbital quantum numbers of the states.   
It is thus quite clear that in the presence of Zeeman coupling, 
we can not use
1) the identification of the two polarized wave amplitudes which
are orthogonal to each other in spin space and propagate independently, 
and 2) the description that the phase effect from the spin degree of freedom 
is effectively some AB effect of a spin-dependent flux. 
We want to point out that the above complexity due to Zeeman coupling
has not been recognized in Ref. \cite{14}.
Consequently the transmission probability obtained therein was wrongly 
simplified by
regarding the contribution from the spin degree of freedom as an
AB effect of some $\mu$-dependent flux for polarized electrons.  
We also want to point out that even when the Zeeman coupling is absent and 
the picture of effective flux is applicable, 
only the geometric phase was included while the
dynamical phase was ignored in the expression of the effective flux 
in Ref. \cite{14}. 

\subsection{Persistent current direction exhibited in transmission
probability}
Numerical calculation has been carried out to illustrate some essential
characteristics of the transmission probability derived here. 
We find that the respective effects of Zeeman coupling and SO interaction 
can be reflected by the resonance of the transmission probability 
in the weak coupling limit at small $\epsilon$. In particular, we can see
an interesting correspondence between the TRS-breaking effects manifested
by the transmission probability and by the persistent current.

We adopt the model of a InAs ring \cite{14}. The Hamiltonian is of the form
\begin{equation}
H_{\rm InAs}=\displaystyle\frac{1}{2m}({\bf p}-\displaystyle\frac{e{\bf A}}  
{c})^{2}+\hbar\kappa [\sigma \!\!\!\!
\sigma\times{\bf p} ]_{z}-\displaystyle\frac{ge\hbar}{4mc}
\sigma\!\!\!\!\sigma\cdot{\bf B},
\end{equation}
where $m=0.023m_{e}$ is the effective mass, $\hbar^{2}\kappa=6.0\times 10^{-10}
{\rm eVcm}$ is the SO coefficient and $g=15$. 
Here the effective electric field is in the z-direction, hence $\chi_{1}=
\pi /2$. For the ring of radius $a=1 
\mu{\rm m}$, the dimensionless coefficient $\alpha$ in Eq. (2) is found to be 
$ma\kappa =1.8$ which is large enough to result in an AC phase of order unity
\cite{15}. The Fermi velocity $v_{F}$ is
approximately $3\times 10^{7} {\rm cms}^{-1}$, corresponding to 
$|n_{F}|\approx 60$.

The effective flux induced by SO interaction
and its effect on the transmission probability can be clearly seen in Fig. 2
where $T_{AB}$ and $\bar{T}_{AB,AC}$ are plotted as functions of $\phi$.
The magnitude of the AC phase can actually 
be approximately measured by a comparison between the 
$\phi$-coordinates of the transmission probabilities' peaks in the absence
and in the presence of the SO interaction. 
The energy splitting due to Zeeman coupling 
is illustrated in Fig. 3. For $\phi=0$ and $\omega_{B}=0$, 
since the Kramers degeneracy makes each two eigenstates of the closed ring 
have the same energy, at certain $E_{F}$ the transmission probabilities in the
two spin branches can reach their highest value 1 simultaneously, thereby 
making $\bar{T}_{AB,AC}=1$.
After the Zeeman coupling is turned on, the resulted energy splitting destroys
the simultaneous happenings of the resonances in the two spin branches and 
we see the maximum values of $T$ decrease with the strength of Zeeman coupling 
appreciably.

With $\epsilon$ being even smaller, the energy dependence of the transmission
probability manifests interesting TRS-breaking effect, which also has its 
corresponding observability in persistent current. In Ref. \cite{17},
it has been demonstrated that in the presence of SO interaction,
the TRS-breaking mechanism due to Zeeman coupling is intrinsically different
from that due to AB flux. As the corresponding observable effect, it has
been found that the direction of the persistent current induced by Zeeman 
coupling changes periodically with the particle number $N$ with the periodicity
$\Delta N=2$ while the direction of the persistent current induced by
AB flux never changes with the particle number. The dependence of the 
current direction on the particle number is actually the dependence 
on Fermi energy. Such energy dependence of the current direction,
an equilibrium phenomenon as it is, can actually be manifested in 
the resonant tunneling of electrons, a transport phenomenon as it is,
in the weak coupling limit. For $\epsilon\rightarrow 0$, the peaks of 
$T(E_{F})$ locate at the eigenenergies $E_{n,\mu}$ of the closed ring 
\cite{18}. In the presence 
of the SO interaction and a weak Zeeman coupling, the transmission
probability is plotted as a function of the incident energy in Fig. 4.
Every two peaks, which are closest to each other, locate at a pair of splitted 
energy levels, which come from the Kramers doublet $(\Psi_{n,\mu},\;
\Psi_{-n-1,-\mu})$ in the absence of Zeeman coupling. 
With the AB flux being zero, the energy splittings in all the splitted
energy levels are the same. Here we use the first-order
perturbation which gives the energy correction but doesn't change the 
eigenfunction. As shown in Ref. \cite{17},
those eigenstates of the closed ring, with increasing energy,
have the spin orientations and current directions in a sequence of 
\begin{equation}\cdot\cdot\cdot ,\;\;[(+,d),(-,u)],\;\;
[(-,d),(+,u)],\;\;[(+,d),(-,u)],\;\;[(-,d),(+,u)],\;\;\cdot\cdot\cdot ,
\end{equation}
where $(s_1,s_2)$ refers to a single quantum state with 
$s_{1}=+$ (anticlockwise) or $-$ (clockwise)
denoting the current direction and $s_{2}=u$ (up) or $d$ (down) denoting the
spin orientation, and $[(s_1,s_2),(-s_1,-s_{2})]$ refers to a pair of 
energy levels from the Kramers doublet.   
The eigenstate correspondence so identified for $T$ 
leads to interesting resonance behavior, as depicted in Fig. 4. It is seen 
that when a small AB flux is added to distinguish the current directions,
each two paired peaks are separated by a distance, which 
takes the larger or the smaller value alternatingly. 
The reason is already clear in the sequence (27). In essence, 
since the current direction determines the sign of energy shift caused by 
a small AB flux,
for $[(-,d),(+,u)]$ the energy splitting due to the small AB flux enhances
that first caused by the Zeeman coupling, while for $[(+,d),(-,u)]$
the energy splitting due to the small AB flux cancels part of that
first caused by the Zeeman coupling.  
We want to point out that the essential character of the above correspondence
between the equilibrium and the transport properties
can be quantitatively, but not be qualitatively, 
affected by the disorder or scattering potential in the ring as long as the
single particle picture holds for electronic states. 
In particular, such correspondence, if experimentally verified or excluded
in some specific ring, 
may serve to detect if the electron-electron interaction
qualitatively alters the electronic states.

\section{Conclusion}
We have studied the motion of electrons confined in 
the perfect ring in the presence of the cylindrically symmetric 
spin-orbit interaction and Zeeman coupling, and the magnetic flux. 
We have obtained the exact solution of the closed ring by using the AA phase
approach in which the geometric and the dynamical phases 
can been explicitly identified for all energy eigenstates.  
Starting from the exact solution for the closed ring, we 
have investigated the transport properties of the ring connected to 
current leads, with emphasis on the roles of the TRS and 
its breaking therein. 
>From the derivation of the transmission probability, we have
shown that in the presence of the TRS-breaking Zeeman coupling,
it is physically impossible to adopt the picture that the spin precession of 
electrons encircling the ring results in some effective, spin-dependent 
Aharonov-Bohm flux in interference, thereby revealing the 
origin of the mistakes in some prior works. 
We have provided the numerical results for illustrating the joint effects of
spin-orbit interaction, Zeeman coupling and magnetic flux. 
>From the resonance behavior of the transmission probability 
in the weak coupling limit, we have 
found the observable correspondence between the TRS-breaking effects 
manifested by the persistent current and 
by the transmission probability as long as the single particle picture
of electronic states holds.

\figure{Schematic representation of the 
electronic waves propagating through the ring 
connected to current leads.
The right junction is located at $\theta =0$ and the left junction at 
$\theta =\pi$ with the upper branch lying within $(0,\;\pi)$ 
and the lower branch within $(\pi,\;2\pi)$.
\label{fig1}}

\figure{Transmission probability as a function of the AB flux for
$\epsilon=0.25$, $ka=60.239$, $a=1\mu{\rm m}$, and 
$\chi_{2}=\displaystyle\frac{\pi}{6}$.
The dotted and the solid lines are assocaited with the absence of 
and the presence of the SO interaction of $\alpha=1.8$ respectively.
\label{fig2}}

\figure{Transmission probability
as a function of the energy of incident electrons ($E_{F}=\hbar^2k^2/2m$), 
for $\epsilon=0.25$, $a=1\mu{\rm m}$, $\alpha=1.8$, 
$\chi_{2}=\displaystyle\frac{\pi}{6}$, and $\phi=0$. The dotted line
corresponds to the absence of SO interaction and Zeeman coupling, 
the solid line corresponds to the presence of SO interaction only, and 
the dashed-dotted line corresponds to the presence of both the 
SO interaction and the Zeeman coupling of $B=30$Gauss. 
\label{fig3}}

\figure{Transmission probability
as a function of the energy of incident electrons ($E_{F}=\hbar^2k^2/2m$), 
for $\epsilon=0.005$, $a=1\mu{\rm m}$, $\alpha=1.8$ and 
$\chi_{2}=\displaystyle\frac{\pi}{6}$.  The solid line
corresponds to the presence of Zeeman coupling of $B=15$Gauss, and
the dashed-dotted line corresponds to 
the presence of the same Zeeman coupling and a magnetic flux of $\phi=0.02$.
a) Constant and alternating distances between paired peaks vs. energy,
represented by the solid and the dashed-dotted lines resepctively.
b) Taken from a) for a clear illustration of the effect caused by 
$\phi=0.02$.
\label{fig4}}

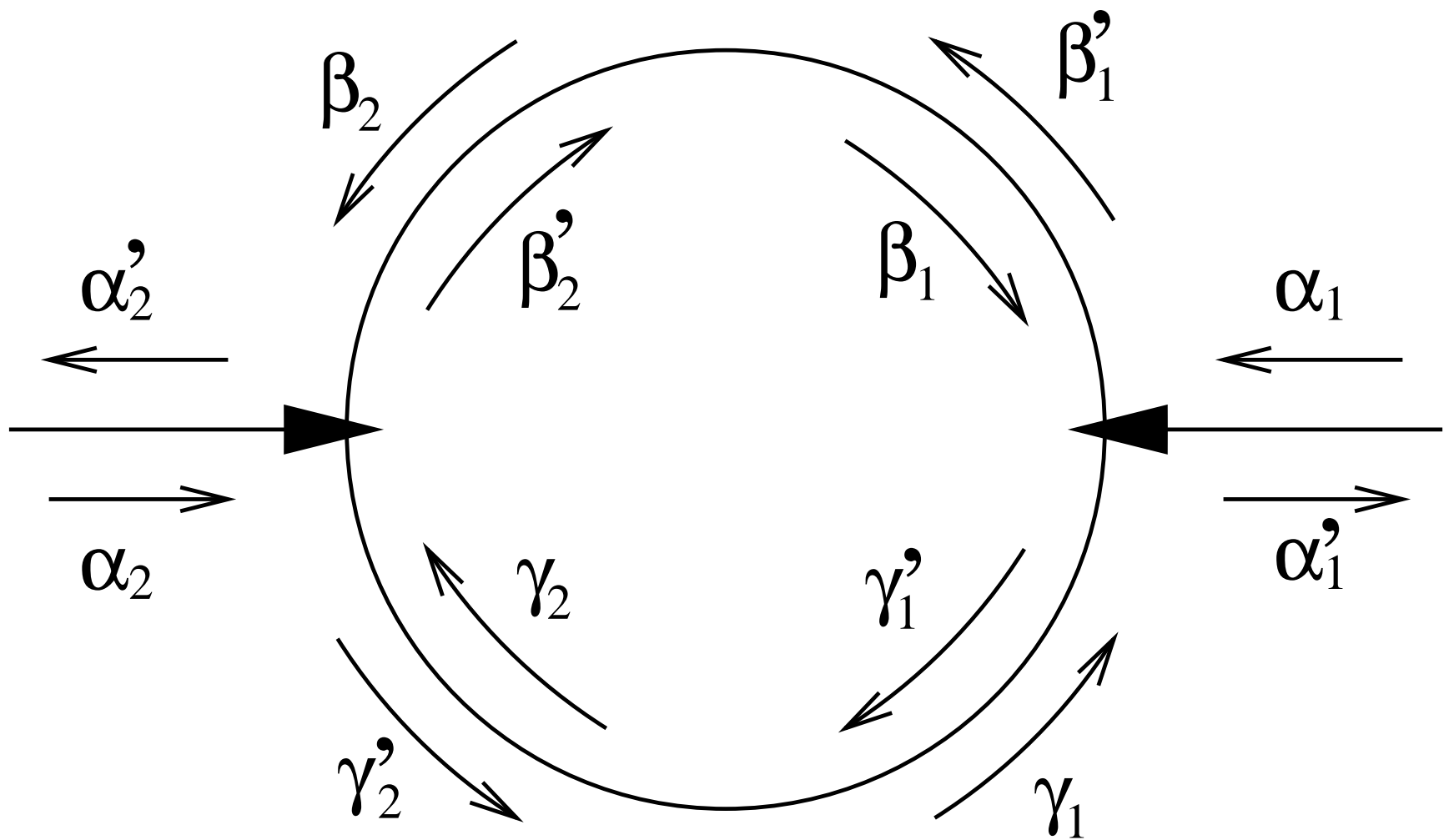
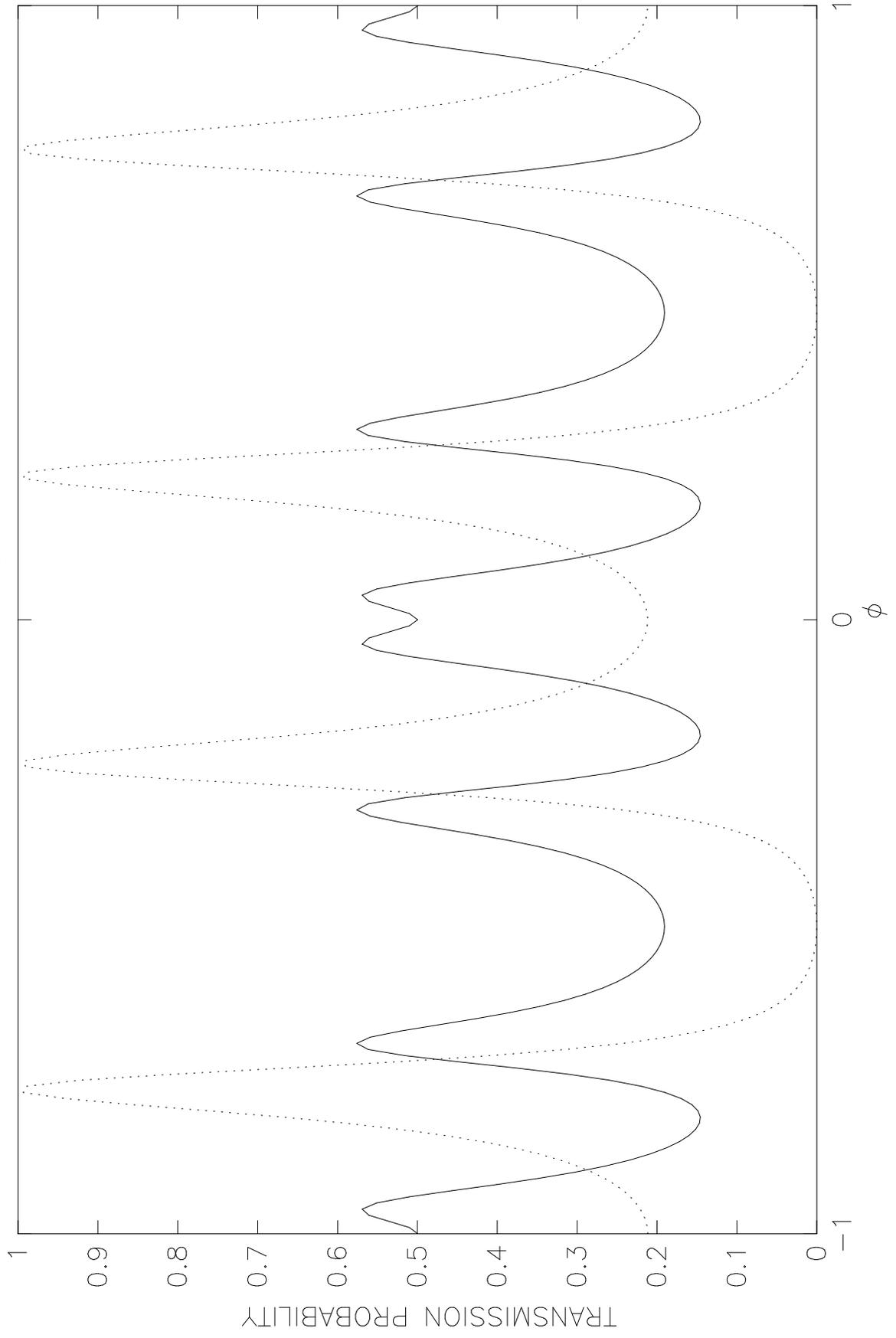
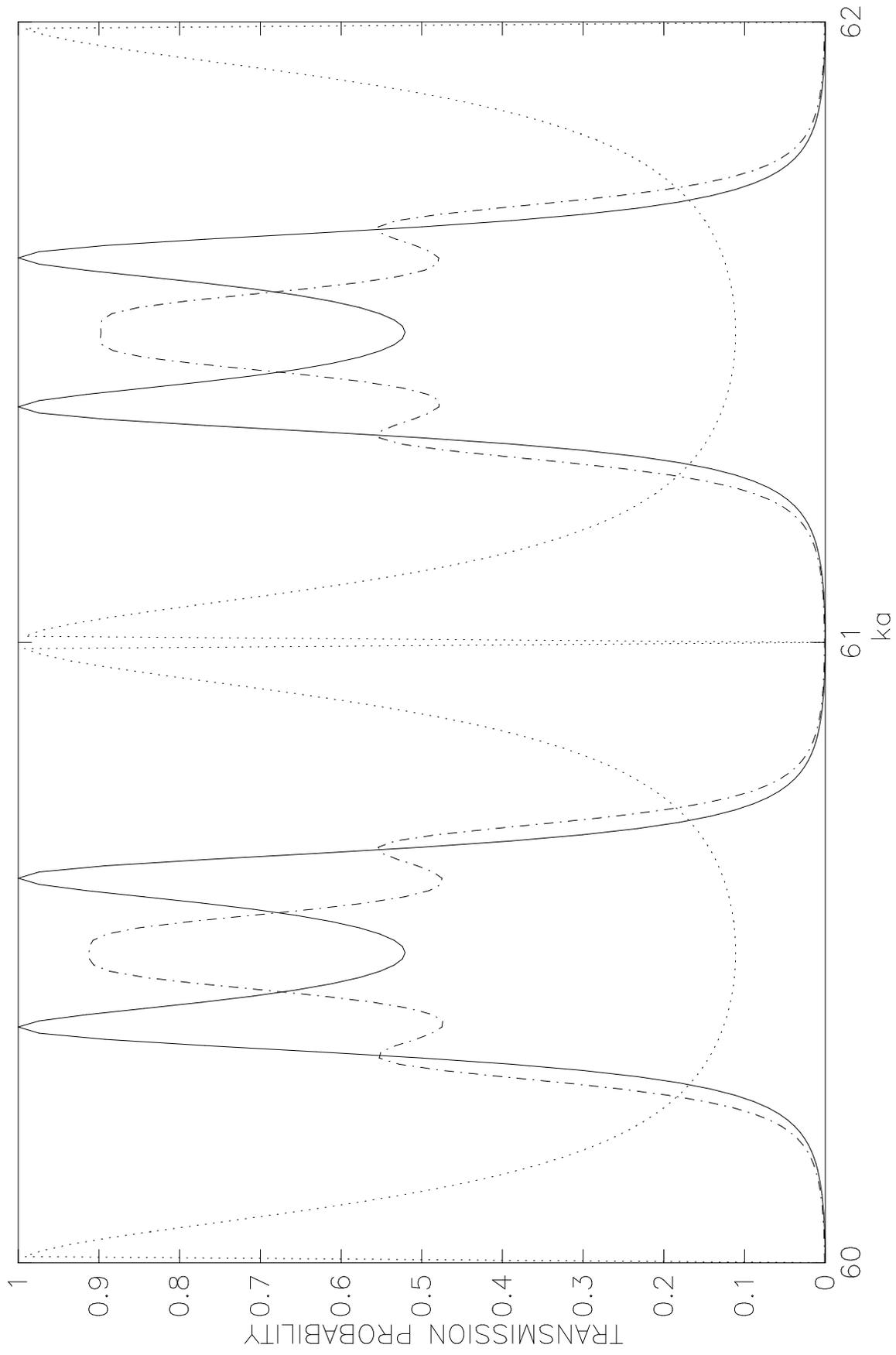
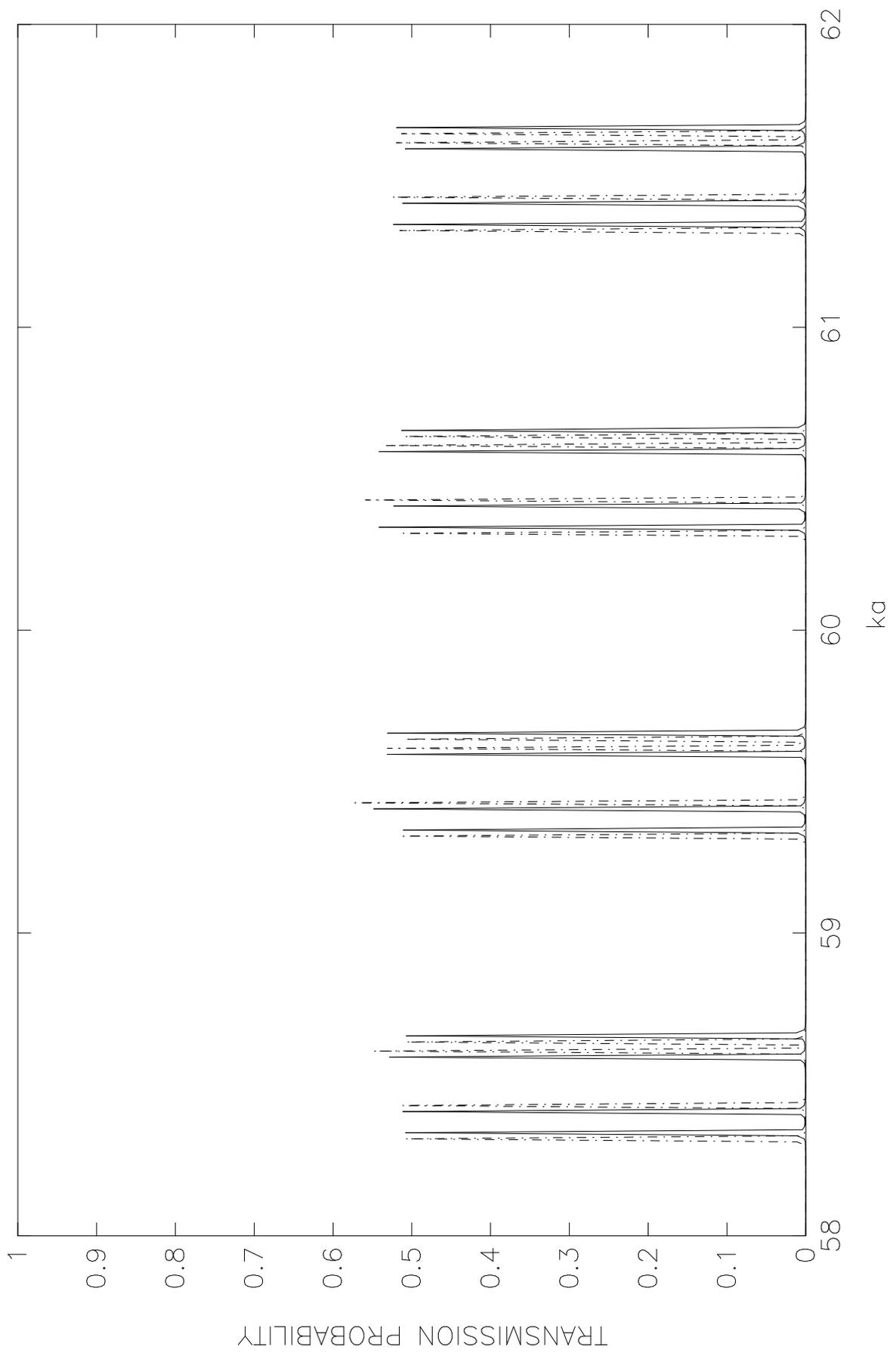
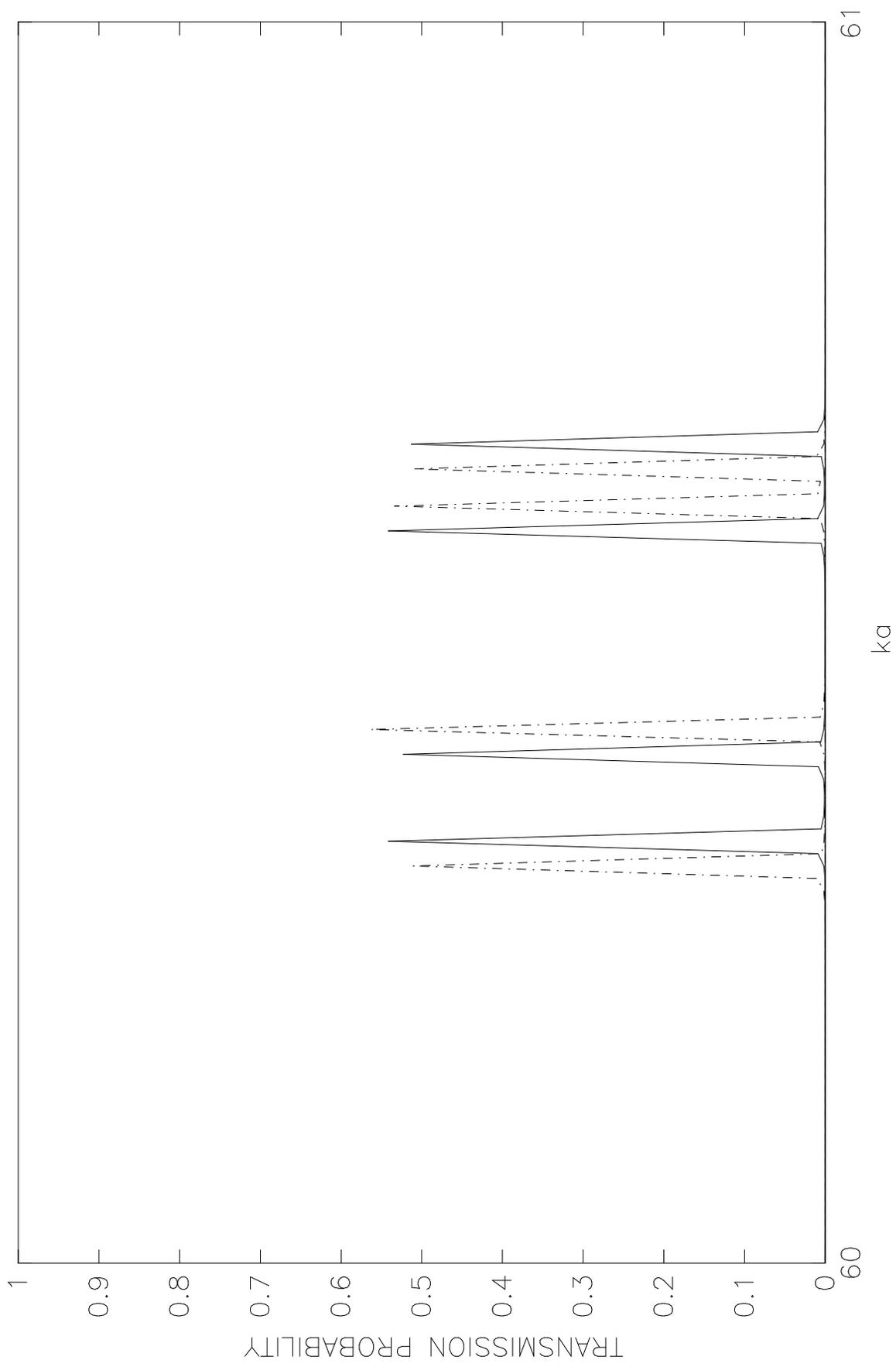
\end{document}